\documentclass[aps, preprint, superscriptaddress]{revtex4-2}
\usepackage{graphicx}
\usepackage{makecell, multirow}
\usepackage{color, bm, soul, xcolor}
\usepackage{amsmath, amssymb}
\usepackage{gensymb,textcomp}
\usepackage{geometry,textcase}
\usepackage{braket}
\usepackage[colorlinks=true, linkcolor=blue, citecolor=blue, urlcolor=blue]{hyperref}
\usepackage{graphicx} 
\usepackage[english]{babel}
\begin{document}

%\title{Percolating Multifractal Domain Inducing Reversible Giant Piezoelectricity}
\title{Percolating Multifractal Domains at a Polymorphic Phase Boundary}
\author{Yuan-Jinsheng Liu}
\thanks{These authors contributed equally}
\affiliation{Fudan University, Shanghai 200433, China}
\affiliation{Department of Physics, School of Science, Westlake University, Hangzhou 310030, China}

\author{Xu Tian}
\thanks{These authors contributed equally}
\affiliation{Department of Physics, School of Science, Westlake University, Hangzhou 310030, China}

\author{Jiahui Zhai}
\thanks{These authors contributed equally}
\affiliation{Department of Physics, School of Science, Westlake University, Hangzhou 310030, China}

\author{Shi Liu}
\email{liushi@westlake.edu.cn}
\affiliation{Department of Physics, School of Science, Westlake University, Hangzhou 310030, China}
\affiliation{Institute of Natural Sciences, Westlake Institute for Advanced Study, Hangzhou 310024, China}

\begin{abstract}
Giant piezoelectricity in ferroelectrics is commonly associated with phase-boundary instabilities, among which the polymorphic phase boundary (PPB) is a prominent example conventionally attributed to the coexistence of ferroelectric phases. Here, using large-scale molecular dynamics simulations of the lead-free (K,Na)NbO$_3$-(Bi,Na)ZrO$_3$ solid solutions, we show that the PPB hosts a percolating multifractal polar domain which governs the dielectric and piezoelectric responses. By quantifying the global fractal dimension and multifractal spectrum width of this polar network, we identify fractal connectivity and multiscale heterogeneity as microstructural order parameters for the PPB. The maximum reversible piezoelectric response occurs when the fractal-domain volume fraction approaches the three-dimensional percolation threshold, suggesting that near-critical polar connectivity enables giant reversible electromechanical coupling. In this mechanism, the fractal backbone preserves polar memory and provides the restoring force required for reversibility, while the surrounding nonfractal regions supply the polar compliance needed for large polarization rotation and strain. These results establish percolating multifractal polar domains as a microscopic mechanism for PPB-enhanced piezoelectricity and suggest fractal connectivity as a design parameter for high-performance piezoelectrics.
\end{abstract}

\maketitle

\clearpage
Giant piezoelectricity in ferroelectric solid solutions has traditionally been rationalized through two types of phase-boundary instabilities~\cite{Li25peadn4926,Rodel09p1153,Fu00p281,Wu15p2559}. The first is the morphotropic phase boundary (MPB), exemplified by lead-based systems such as Pb(Zr,Ti)O$_3$ and Pb(Mg$_{1/3}$Nb$_{2/3}$)O$_3$--PbTiO$_3$, where two or more ferroelectric phases with different polarization symmetries become degenerate at a nearly vertical boundary in the composition--temperature phase diagram~\cite{JAFFE71,Noheda99p2059,Ahart08p545,Park97p1804,Damjanovic98p1267}. Because the MPB is primarily controlled by composition and exhibits only weak temperature dependence, it provides a broad operating window~\cite{Li10p252903}. The second route is the polymorphic phase boundary (PPB), widely exploited in lead-free (K,Na)NbO$_3$-based materials, where chemical substitution shifts an intrinsically temperature-driven ferroelectric--ferroelectric transition toward the desired operating temperature~\cite{Zhang07p251,Li13p3677,Lv20p224,Wu15p2559}. Unlike an MPB, a PPB is a strongly temperature-dependent instability whose functional enhancement is confined to a narrow thermal interval~\cite{Lv25p859,Liu09p257602}. 
%Although optimized PPB compositions can approach lead-based piezoelectrics in performance, it remains unclear how the temperature-dependent phase instability leads to large field-induced deformation while preserving reversibility, meaning that the polarization and strain largely return to their original state after the electric field is removed. Resolving this mechanism is essential for developing environmentally friendlier lead-free piezoelectrics.

The PPB exhibits signatures that cannot be understood by simply extending the conventional MPB picture. In both MPB and PPB systems, enhanced piezoelectricity is usually attributed to the coexistence of nearly degenerate ferroelectric phases with different polarization directions, which lowers the barrier for field-induced polarization rotation or phase conversion~\cite{Noheda99p2059,Fu00p281,Ahart08p545,Wu15p2559,Ochoa16p142901}. In lead-based MPB compositions, this phase instability is typically accompanied by enhanced dielectric susceptibility, so that the piezoelectric and dielectric responses peak in the same region~\cite{Wang12p433}. By contrast, PPB systems often show a broad dielectric plateau rather than a sharp anomaly, and the piezoelectric coefficient reaches its maximum at $T^*$ within this plateau~\cite{Garcia20p131102}. Upon heating beyond $T^*$, the piezoelectric response decreases rapidly. This dielectric–piezoelectric decoupling is difficult to reconcile with a simple flattened-energy-landscape picture, in which easier polarization rotation would be expected to enhance both dielectric and electromechanical responses~\cite{Fu00p281,Lv25p859,Wang16p627}. Reversibility imposes an additional constraint: after field removal, the polarization and strain must return largely to their original state. Nearly degenerate polar states facilitate field-driven rotation, but they can also promote trapping in a different local minimum. A reversible piezoelectric response therefore requires a restoring mechanism that preserves memory of the original polar state. Resolving this microscopic mechanism is essential for developing environmentally friendlier lead-free piezoelectrics.

In this work, we use large-scale molecular dynamics (MD) simulations to investigate temperature-dependent polar microstructure of (K,Na)NbO$_3$--(Bi,Na)ZrO$_3$ (KNN--BNZ), a model lead-free solid-solution system known for PPB-enhanced piezoelectricity~\cite{Lv25p859}. We find that the PPB hosts a percolating multifractal network: a connected but spatially heterogeneous polar structure consisting of dense clusters and tenuous branches across multiple length scales. This heterogeneity is quantified by the multifractal spectrum width \(\Delta\alpha\), with larger \(\Delta\alpha\) indicating greater variation in the local spatial scaling of polar regions. The PPB in KNN-BNZ coincides with the temperature window where the multifractal domains exhibit anomalously large $\Delta\alpha$. Importantly, this connected multifractal structure provides a microscopic mechanism for giant reversible piezoelectricity. The fractal domain acts as a backbone that preserves memory of the initial polar state and provides a restoring force after field removal, while the surrounding nonfractal regions contain flexible dipoles that rotate readily under an applied field and generate large strain. Upon further heating, this backbone collapses, as indicated by a decrease in the global fractal dimension. This collapse coincides with a continued increase in dielectric response but a reduction in piezoelectricity. We further find that the piezoelectric response is maximized when the fractal-domain volume fraction is \(\omega_f\approx 0.371\), close to the three-dimensional percolation threshold~\cite{Lorenz98p8147}. These results suggest that near-threshold connectivity of a multifractal polar domain is a microscopic origin of giant reversible electromechanical coupling at the PPB.

\begin{figure}[t]
\includegraphics[width=1.0\textwidth]{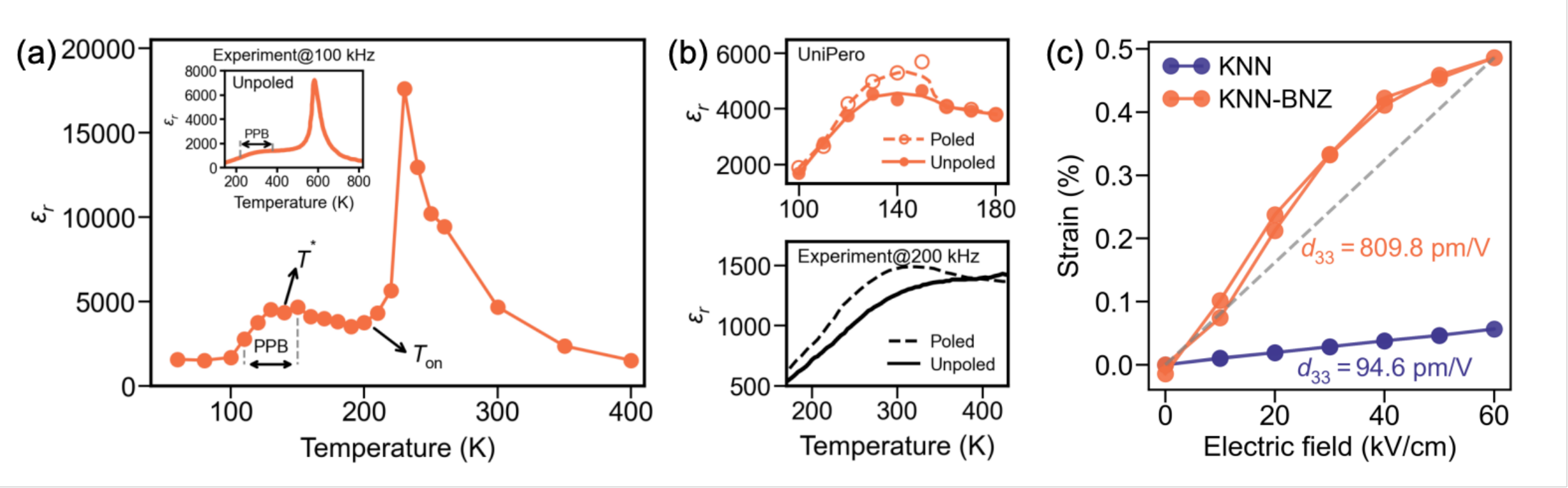}
\centering
\caption{MD simulations of the polymorphic phase boundary in lead-free KNN-BNZ solid solutions.
(a) Simulated temperature-dependent relative dielectric permittivity, \(\epsilon_r\), of
0.94(K\(_{0.48}\)Na\(_{0.52}\))NbO\(_3\)--0.06(Bi\(_{0.5}\)Na\(_{0.5}\))ZrO\(_3\)
(KNN-BNZ), compared with the experimental dielectric spectrum shown in the inset.
The simulation reproduces key features associated with the polymorphic phase boundary (PPB),
including a local maximum around \(T^* \approx 140~\mathrm{K}\), a dielectric plateau over
\(110\)--\(150~\mathrm{K}\), and a sharp increase above
\(T_{\mathrm{on}} \approx 200~\mathrm{K}\).
(b) Dielectric response of poled and unpoled samples near the PPB. The top panel shows
MD simulations performed using UniPero, while the bottom panel presents
experimental measurements from Ref.~\cite{Lv25p859}. The simulations capture the subtle differences
between the poled and unpoled states.
(c) Simulated strain--electric-field loops at \(T^* = 140~\mathrm{K}\) for KNN--BNZ and pure KNN.
The slope gives an effective piezoelectric coefficient, \(d_{33}\), of
\(809.8~\mathrm{pm/V}\) for KNN-BNZ and \(94.6~\mathrm{pm/V}\) for pure KNN, demonstrating the
nearly order-of-magnitude enhancement in piezoelectric response associated with the PPB.
}
\label{fig:dielectric}
\end{figure}

We first validate the MD description of the PPB based on the temperature-dependent dielectric response. The simulations employ UniPero, a machine-learning force field developed in our previous work~\cite{Wu23p180104}. UniPero is applicable to a broad chemical space of perovskite oxides and has been extensively benchmarked for various ferroelectric perovskites~\cite{Hu24p046802,Liu26p021022}. As a baseline, UniPero reproduces temperature-driven orthorhombic--tetragonal--cubic phase sequence of ferroelectric KNN [Figs.~S1(h) and S2]~\cite{Li13p3677}. 
With only 6\% BNZ substitution, 0.94(K$_{0.48}$Na$_{0.52}$)NbO$_3$--0.06(Bi$_{0.5}$Na$_{0.5}$)ZrO$_3$ exhibits a PPB. Experimentally, this PPB is marked by a broad dielectric plateau over a finite temperature window, followed by a sharp increase toward the main dielectric peak [inset of Fig.~\ref{fig:dielectric}(a)]~\cite{Lv25p859}. Remarkably, our MD simulations of KNN--NBZ at the same composition reproduce the main features of this temperature-dependent dielectric spectrum [Fig.~\ref{fig:dielectric}(a)]. Specifically, the simulated spectrum exhibits a local maximum around $T^*=140$~K associated with a dielectric plateau from 110 to 150~K, followed by a sharp rise in the dielectric response beginning at $T_{\rm on}=200$~K. The simulations also capture the subtle differences between poled and unpoled samples near the PPB region [Fig.~\ref{fig:dielectric}(b)]. Additional validation tests, including comparisons of KNN and KNN--BNZ configuration energies with \textit{ab initio} MD results at different temperatures, further confirm the accuracy of UniPero (see Figs.~S4--S5).

The validity of UniPero for modeling PPB-enhanced electromechanical response is also demonstrated by the simulated strain--electric-field loop [Fig.~\ref{fig:dielectric}(c)]. From the slope of the curve, we extract an effective piezoelectric coefficient $d_{33}$ (see additional discussions below). At $T^*=140$~K within the PPB region, KNN--BNZ exhibits a large $d_{33}=809.8$~pm/V, nearly an order of magnitude larger than that of pure KNN at the same temperature ($d_{33}=94.6$~pm/V). This strong enhancement is consistent with the experimentally known PPB-enhanced piezoelectricity. 
Additionally, the magnitude of $d_{33}$ for KNN-BNZ decreases rapidly as the temperature rises beyond the PPB region (Fig.~S3),  reproducing the experimentally observed decoupling between the dielectric and piezoelectric peaks~\cite{Wang16p627}. 
We note that MD simulations underestimate the absolute phase-transition temperatures, a common issue in ferroelectric simulations~\cite{Wu23p144102,Wu23p180104,Shi24p174104,Liu13p104102,Gindele15p17784}. This discrepancy likely originates from the exchange-correlation functional used to train the force field, such as PBEsol~\cite{Perdew08p136406}, as suggested by a recent benchmark study on Curie temperature~\cite{Li26p}. Nevertheless, the reproduced dielectric plateau, onset of the main dielectric rise, poled--unpoled differences, and giant piezoelectric enhancement indicate that the simulations capture the key physics of the PPB. The comparison with experiments can therefore be understood through a temperature rescaling.

All MD simulations are performed in the isobaric--isothermal (\textit{NPT}) ensemble using \texttt{LAMMPS}~\cite{Plimpton95p1}, with a 2~fs time step, a Nos\'e--Hoover thermostat, and a Parrinello--Rahman barostat. The KNN--BNZ system is modeled by a \(32 \times 32 \times 32\) supercell containing 163,840 atoms and is equilibrated for 1~ns at each temperature before data collection (see Supplemental Sects.~II.A--II.B).
To analyze the polar microstructure at the PPB, we develop a domain-growing method to identify statistical polar domains. Specifically, a unit cell with a robust local polarization (defined as the dipole moment per unit-cell volume) is randomly selected as a seed. Neighboring cells are incorporated into the growing domain only if they also exhibit sizable local polarization and their polarization directions remain within $30^\circ$ of the domain polarization direction over a 200-ps correlation time. This procedure is repeated iteratively for the updated domain and its neighboring cells until no additional cells satisfy the growth criteria (see Supplementary Sect.~II.C).

\begin{figure}[t]
\includegraphics[width=1.0\textwidth]{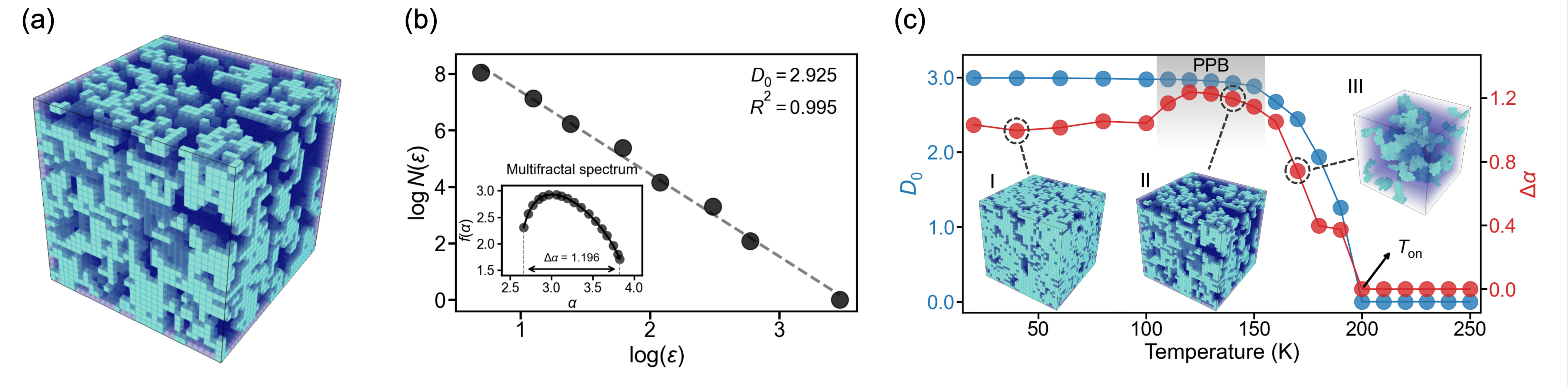}
\centering
\caption{
Polar multifractal domains at the PPB.
(a) Polar multifractal domain identified at \(T^* = 140~\mathrm{K}\), forming a single connected network that percolates through the simulation cell.
(b) Box-counting analysis of the domain shown in (a), confirming fractal scaling with a global fractal dimension of \(D_0 = 2.925\). The inset shows the multifractal spectrum, \(f(\alpha)\), as a function of the local scaling exponent, \(\alpha\). The broad spectrum width, \(\Delta\alpha = 1.196\), together with its right-skewed profile, indicates a heterogeneous domain structure composed of a dense backbone and sparse, weakly connected regions.
(c) Temperature evolution of \(D_0\) and \(\Delta\alpha\), with representative fractal domain configurations at \(40~\mathrm{K}\) (I), \(140~\mathrm{K}\) (II), and \(170~\mathrm{K}\) (III). The multifractal width \(\Delta\alpha\) reaches a maximum within the PPB plateau, \(110\)--\(150~\mathrm{K}\), suggesting that it serves as a microstructural order parameter for the PPB. Beyond the PPB, both \(D_0\) and \(\Delta\alpha\) decrease sharply and vanish near \(T_{\mathrm{on}} \approx 200~\mathrm{K}\), where the fractal domain disintegrates into isolated polar clusters (denoted as $D_0=0$).
}
\label{fig:fractal}
\end{figure}

At $T^*=140$\,K within the PPB region, the identified polar domain forms a connected, system-spanning network [Fig.~\ref{fig:fractal}(a)].  We quantify this domain using box-counting analysis, where the simulation supercell is divided into boxes of size \(\epsilon\), and a box is counted as occupied if it contains at least one unit cell belonging to the identified polar domain. The number of occupied boxes, \(N(\epsilon)\), is then measured as \(\epsilon\) varies. The nearly linear relation between $\log N(\epsilon)$ and $\log(\epsilon)$ confirms fractal scaling of this polar network [Fig.~\ref{fig:fractal}(b)]~\cite{Mandelbrot67p636,Feder1988,Catalan08p027602}. We therefore refer to this system-spanning polar network as a fractal domain. From the slope of the box-counting relation, we obtain the global fractal dimension $D_0=2.925$, indicating a nearly space-filling but noncompact network at the PPB.

Beyond its global fractal scaling, the PPB fractal domain also exhibits strong spatial heterogeneity. We quantify this heterogeneity using the multifractal spectrum, where \(\alpha\) denotes the local scaling exponent and \(f(\alpha)\) gives the fractal dimension of the polar regions characterized by \(\alpha\) (see Supplementary Sect.~III)~\cite{Halsey86p112,Halsey86p1141}. If all parts of the domain had similar density and connectivity, the spectrum width would be $\Delta\alpha=0$; a larger $\Delta\alpha$ indicates a broader distribution of local environments (see Fig.~S6 for an example). A larger $f(\alpha)$ also indicates that regions with the corresponding $\alpha$ are more spatially extensive. 
Our calculations reveal that the spectrum is broad at 140~K, with $\Delta\alpha=1.196$ [inset of Fig.~\ref{fig:fractal}(b)]. The spectrum is also skewed toward lower \(\alpha\), indicating that dense regions dominate the multifractal domain, with sparse, weakly connected regions also present.

Figure~\ref{fig:fractal}(c) shows the temperature evolution of the global fractal dimension $D_0$ and multifractal spectrum width $\Delta\alpha$, together with representative domain structures at 40~K (I), 140~K (II), and 170~K (III). Notably, $\Delta\alpha$ acts as an effective microstructural order parameter for the PPB. At low temperature, represented by 40~K, $D_0$ is close to 3 and $\Delta\alpha\approx 1$, indicating a nearly space-filling domain with moderate heterogeneity. Upon entering the PPB window ($\approx$110--150~K), represented by $T^*=140$~K, $D_0$ remains high, while $\Delta\alpha$ reaches its maximum, identifying the PPB as a regime characterized by a highly heterogeneous multifractal domain. Such a domain contains local polar regions with diverse environments, potentially enabling a broad range of local dipole-reorientation modes under an electric field and thereby producing a local dielectric peak.
Beyond the PPB, represented by 170~K, the fractal domain begins to disintegrate: $D_0$ drops steeply, reaching $\approx 1.94$ at 180~K and $\approx 1.26$ at 190~K, while $\Delta\alpha$ falls below 0.40. At 200~K and above, both $D_0$ and $\Delta\alpha$ vanish, corresponding to the collapse of the fractal domain into isolated polar clusters. This collapse coincides with $T_{\rm on}$, where the dielectric constant leaves the plateau and rises toward the main peak [Fig.~\ref{fig:dielectric}(a)].

We suggest that the dielectric plateau before $T_{\rm on}$ could arise from the coupled evolution of $D_0$ and $\Delta\alpha$. As temperature increases beyond the PPB, the reduction in $\Delta\alpha$ indicates diminished domain heterogeneity, which may weaken the contribution from diverse local dipole-reorientation modes associated with regions of different polar densities. Meanwhile, the decrease in $D_0$ indicates that the fractal domain becomes less space-filling, leaving more weakly constrained non-domain dipoles that can be more easily reoriented by an electric field. These two effects may partially compensate each other, leading to a weakly temperature-dependent dielectric response. 
%Thus, the coupled temperature evolution of $D_0$ and $\Delta\alpha$ provides a plausible microstructural basis for the complex dielectric behavior of KNN-BNZ solid solutions.

\begin{figure}[t]
\includegraphics[width=1.0\textwidth]{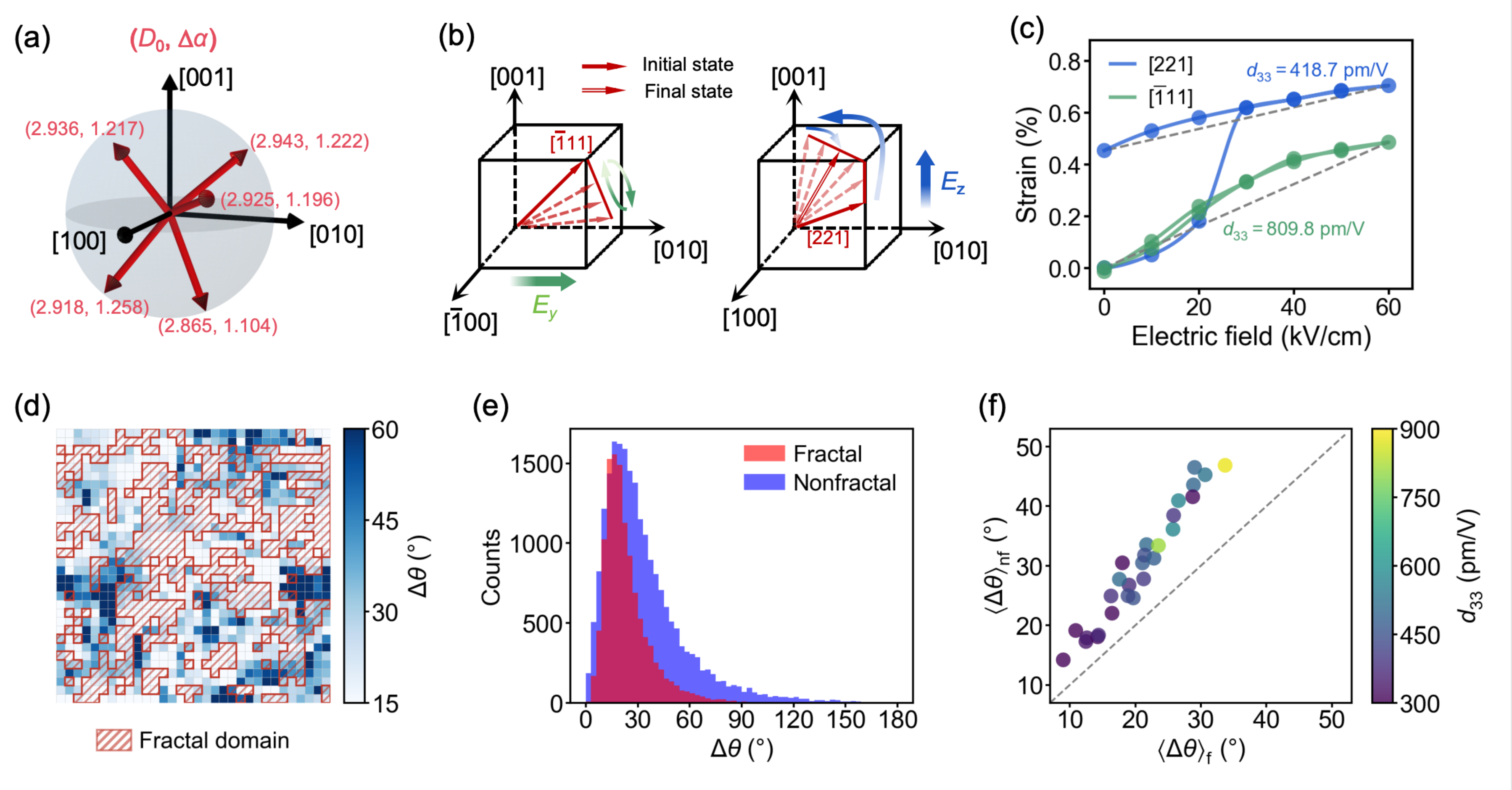}
\centering
\caption{
Fractal-domain-assisted reversible giant piezoelectricity at the PPB.
(a) Representative stable orientations of the total net polarization of the supercell obtained from MD simulations at \(T^*\). Each polarization state is associated with a polar fractal domain characterized by distinct \(D_0\) and \(\Delta\alpha\) values, as labeled.
(b) Schematic of field-driven polarization rotation for two representative states: the \([\bar{1}11]\) state driven by \(E_y\) and the \([221]\) state driven by \(E_z\). The electric field is gradually increased to \(E_{\mathrm{max}} = 60~\mathrm{kV/cm}\) and then removed. The \([\bar{1}11]\) state follows a reversible path and returns to its initial orientation, whereas the \([221]\) state rotates irreversibly and becomes trapped near \([114]\).
(c) Corresponding strain--electric-field loops for the two rotation paths in (b).
(d) Spatial map of unit-cell-resolved local polarization rotation angles, \(\Delta\theta\), in a representative supercell layer at \(E_{\mathrm{max}}\). Hatched regions denote the polar fractal domain.
(e) Distributions of \(\Delta\theta\) for all unit cells inside and outside the fractal domain. Dipoles within the fractal domain show a narrower distribution centered at smaller rotation angles, whereas those outside the domain exhibit a broader distribution extending to larger rotations.
(f) Correlation between the average dipole rotation angles inside and outside the fractal domain at $E_{\rm max}$, \(\langle \Delta\theta \rangle_{\mathrm{f}}\) and \(\langle \Delta\theta \rangle_{\mathrm{nf}}\), respectively, for more than 20 initial polar states at \(T^*\). In all cases, \(\langle \Delta\theta \rangle_{\mathrm{nf}}\) exceeds \(\langle \Delta\theta \rangle_{\mathrm{f}}\), indicating that the rigid fractal domain acts as a restoring backbone, while the nonfractal region provides the compliance required for large polarization rotation and strain.
}
\label{fig:electric_field}
\end{figure}

We next examine the connection between the multifractal domain and PPB-enhanced piezoelectricity. At $T^*=140$~K, MD simulations show that the free-energy surface is highly flexible with respect to the global polarization direction. By preparing KNN-BNZ supercells with different initial poling directions and then equilibrating them at zero field, we find that the global polar state can stabilize along more than 50 distinct directions. Each stable state is associated with a distinct multifractal domain characterized by its own \(D_0\) and \(\Delta\alpha\). Representative polarization orientations are shown in Fig.~\ref{fig:electric_field}(a), including states close to \([\bar{1}11]\) and more strongly tilted states such as \([221]\). This multiplicity of stable global polar states indicates a highly degenerate energy landscape at the PPB, making the origin of reversible piezoelectricity nontrivial.

Taking the initial global polar state oriented along \([\bar{1}11]\) as an example, we simulate the strain--electric-field response under a cyclic field. To mimic the experimental measurement, an electric field is applied along the Cartesian $y$-axis ($[010]$ direction), gradually increased to $E_{\rm max}=60$~kV/cm, and then reduced to zero [left panel of Fig.~\ref{fig:electric_field}(b)]. During this cycle, the polarization rotates reversibly and produces a closed strain--electric-field loop [Fig.~\ref{fig:electric_field}(c)]. Following the experimental convention, the effective coefficient is denoted as $d_{33}$ because it is extracted from the strain change along the field direction; its value exceeds 800~pm/V. Analysis of the local polarization response shows that this giant reversible response arises from the dual role of the fractal domain: it serves as a relatively rigid backbone that retains the initial polar state, while the surrounding nonfractal regions are more compliant and rotate more readily under the field, generating large strain. After field removal, the backbone drives the system back to its initial polar configuration. Figure~\ref{fig:electric_field}(d) shows a contour map of the dipole rotation angles in one layer of the supercell at $E_{\rm max}$, confirming that dipoles within the fractal domain rotate much less than those in the nonfractal regions (see all layers in Figs.~S7-S8). Correspondingly, the \(\Delta\theta\) distributions in Fig.~\ref{fig:electric_field}(e) show that the fractal domain has a narrow distribution concentrated at small rotation angles, whereas the nonfractal region exhibits a broad distribution extending to much larger angles.

A large polarization-rotation angle is expected to generate a large strain change and is therefore important for a strong piezoelectric response. However, the presence of a fractal domain alone does not guarantee reversibility: the field-induced rotation must avoid other stable global polar states. This condition is illustrated by the initial $[221]$ state [right panel of Fig.~\ref{fig:electric_field}(b)]. When an electric field is applied along $[001]$, the system develops a large strain, but the rotation path passes near another global polar state oriented along $[114]$.  After field removal, the system becomes trapped in this new state rather than returning to the initial \([221]\) state. As a result, the strain--electric-field loop is open [Fig.~\ref{fig:electric_field}(c)], and subsequent application of the same field activates only a smaller piezoelectric response of \(d_{33}=418.7\)~pm/V. %The evolution of $D_0$ confirms after the field cycle, $D_0$ does not recover its initial value, indicating that the original fractal backbone has been reconfigured.

The contrast between \([\bar{1}11]\) and \([221]\) states establishes two requirements for giant reversible piezoelectricity at the PPB. First, the system must contain a fractal backbone that is sufficiently rigid to preserve the initial polar state and provide a restoring force. Second, the applied field must rotate the polarization along a path that avoids capture by another stable fractal-domain configuration. These requirements are consistent with experiments on KNN-based ceramics, where proper poling is often essential for optimizing piezoelectric properties~\cite{Zheng16p9242,Liu23p4044}. Pre-poling selects a favorable initial polar state, while an appropriate driving field activates a reversible polarization-rotation pathway. Depending on the poling treatment, the same composition can exhibit either a closed strain--electric-field loop with a large reversible response or an open loop with a reduced response upon subsequent cycling~\cite{Lv25p859,Wu23p1009,Huan16p22053,Wu24p2408}.

To further test the role of the fractal domain more broadly, we analyzed the reversible piezoelectric response of more than 20 zero-field global polar states at $T^*$. For each state, we chose the field direction so that polarization rotation remained reversible at a fixed maximum field of \(E_{\rm max}=60\)~kV/cm, producing a closed strain--electric-field loop from which \(d_{33}\) was extracted. Because all accessible zero-field polar states are known, the field-induced rotation path can be selected to avoid trapping by other global states. For the reversible polarization-rotation path of each state, we calculated the average dipole rotation angles at \(E_{\rm max}\) inside and outside the fractal domain, denoted as \(\langle\Delta\theta\rangle_{\rm f}\) and \(\langle\Delta\theta\rangle_{\rm nf}\), respectively. As shown in Fig.~\ref{fig:electric_field}(f), $\langle\Delta\theta\rangle_{\rm nf}$ is consistently larger than $\langle\Delta\theta\rangle_{\rm f}$, confirming that nonfractal dipoles are more field-responsive, whereas the fractal domains remain comparatively rigid. The magnitude of $d_{33}$ generally increases with the rotation amplitude, especially with $\langle\Delta\theta\rangle_{\rm nf}$. These results demonstrate the intimate connection between the fractal domain and giant reversible piezoelectricity in KNN-BNZ.

\begin{figure}[t]
\includegraphics[width=0.6\textwidth]{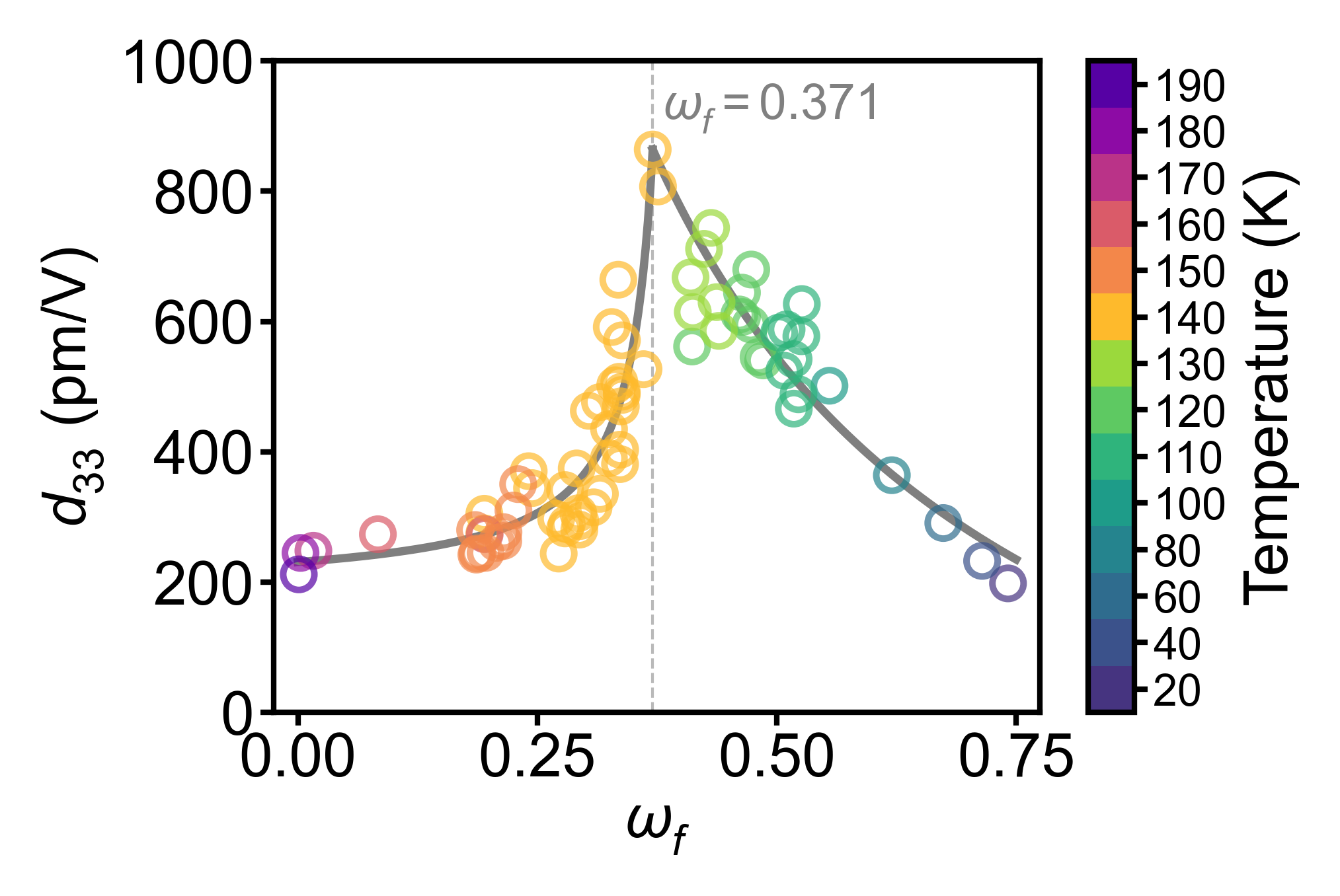}
\centering
\caption{
Volcano plot of \(d_{33}\) as a function of the fractal-domain volume fraction, \(\omega_f\).
The piezoelectric coefficients are computed for various polarization states over a range of temperatures.
The volcano-shaped trend peaks at \(\omega_f \approx 0.371\), where \(d_{33}\) exceeds \(800~\mathrm{pm/V}\).
This optimal volume fraction, realized at \(T^*\) within the PPB, also lies near the critical threshold of 0.312 predicted by three-dimensional percolation theory~\cite{Lorenz98p8147}.
}
\label{fig:volcano}
\end{figure}

The preceding analysis shows that, even at a fixed temperature, different zero-field polar states can yield a distribution of reversible piezoelectric coefficients. This complexity is further increased by the intrinsic temperature dependence of $d_{33}$. Remarkably, both the polar-state and temperature dependences of $d_{33}$ can be captured by a single microstructural parameter: the fractal-domain volume fraction, $\omega_f$. At each temperature, we sampled multiple metastable fractal-domain configurations (global polar states), determined their $\omega_f$, and calculated the reversible \(d_{33}\) from strain--electric-field loops with \(E_{\rm max}=60\)~kV/cm. Plotting all data together as $d_{33}$ versus $\omega_f$ reveals a clear volcano-shaped trend (Fig.~\ref{fig:volcano}). The maximum response, exceeding 800~pm/V, occurs at $\omega_f \approx 0.371$, which is reached within the PPB temperature region and is markedly close to the critical threshold of three-dimensional percolation theory, approximately 0.312 for random isotropic systems~\cite{Lorenz98p8147}. This suggests that the largest reversible piezoelectricity emerges when the fractal domain in the PPB operates near percolation criticality.

A heuristic interpretation of the optimal \(\omega_f\) follows from the dual role of the fractal domain. When $\omega_f$ is too small, the fractal backbone is weakly connected and provides only a weak restoring force. In this case, the system can sustain only a small electric field for reversible rotation, limiting the maximum rotation angle and strain change. When $\omega_f$ is too large, the backbone is sufficiently rigid, but the volume of soft nonfractal regions becomes too small to accommodate large polarization rotation and strain. It is possible that near the percolation threshold when the structure is ``just rigid, maximally soft'', the connected fractal backbone provides enough restoring force for reversible response under a large driving field, while the remaining nonfractal matrix supplies the compliance needed for large strain. In this picture, the temperature dependence of \(d_{33}\) is primarily governed by the evolution of \(\omega_f\), with the PPB region providing the critical fractal-domain fraction that maximizes the piezoelectric response. %This suggests a design principle: tune composition so that the optimal $\omega_f$ occurs at the target operating temperature.

In summary, our results identify the PPB as a percolating multifractal polar domain with strong spatial heterogeneity. The global fractal dimension \(D_0\) and multifractal spectrum width \(\Delta\alpha\) serve as microstructural order parameters that rationalize the temperature-dependent dielectric response, including the dielectric plateau. Within this framework, PPB-enhanced giant piezoelectricity originates from the dual role of the fractal domain: the connected backbone provides the restoring force required for reversibility, while the surrounding nonfractal regions supply the compliance needed for large strain. The maximum reversible \(d_{33}\) occurs when the fractal-domain volume fraction \(\omega_f\) reaches its optimal value in the PPB region, close to the three-dimensional percolation threshold. These findings suggest that tuning fractal connectivity toward percolation criticality may guide the design of reversible deformation in other structurally disordered systems.

\clearpage
{\bf{Acknowledgments}} We acknowledge the support from National Natural Science Foundation of China (92370104) and Zhejiang Provincial Natural Science Foundation of China (LR25A040004). The computational resource is
provided by the Open Source Supercomputing Center of
S-A-I. 

\clearpage
\bibliography{SL}

\clearpage

\clearpage

\clearpage

\clearpage

\clearpage
\end{document}